\documentclass[onecollarge,natbib]{svjour2}
\bibpunct{[}{]}{;}{n}{}{,} 
\smartqed  
\usepackage{graphicx}
\journalname{Few-Body Systems}

\newcommand{\blf}[1]{\bf  {\tilde #1}}
\newcommand{\bq}{\begin{eqnarray}}
\newcommand{\eq}{\end{eqnarray}}
\newcommand{\bm}[1] {\mbox{\boldmath{$#1$}}}
\def\Tr{\rm Tr}
\begin{document}
\title{A Light-Front approach to the $^3$He spectral function}
%
%

\author{Sergio Scopetta \and
        Alessio Del Dotto \and
        Leonid Kaptari \and
Emanuele Pace \and
	    Matteo Rinaldi \and
	    Giovanni Salm\`e 
}

\institute{S. Scopetta \at
 Dipartimento di Fisica e Geologia, Universit\`a di Perugia and INFN, 
Sezione di Perugia, Italy \\
             Tel.: +39-075-5852721\\
              Fax: +39-075-44666\\
              \email{sergio.scopetta@pg.infn.it}
\and
A. Del Dotto \at
   Dipartimento di Fisica,        
Universit\`a di Roma Tre and INFN, Roma 3, Italy
\and
L. Kaptari \at
Bogoliubov Lab. Theor. Phys., 141980, JINR, Dubna, Russia
\and
E. Pace \at
 Dipartimento di Fisica, Universit\`a di Roma ``Tor Vergata'' and INFN, Roma 2, Italy
\and
M. Rinaldi \at
 Dipartimento di Fisica e Geologia, Universit\`a di Perugia and INFN, 
Sezione di Perugia, Italy
	   \and
G. Salm\`e \at
           INFN Sezione di Roma, Italy
          }
\date{Received: date / Accepted: date}

\maketitle

\begin{abstract}
The analysis of semi-inclusive deep inelastic electron scattering 
off polarized $^3$He 
at finite  momentum transfers, aimed at 
the extraction of the quark transverse-momentum 
distributions in the neutron, requires the use of
a distorted spin-dependent spectral function for $^3$He, which  
takes care of the final state
interaction effects. This quantity is introduced in the non-relativistic case, 
and its generalization in a Poincar\'e
covariant framework, in plane wave impulse approximation for the moment 
being, is outlined.
Studying the light-front spin-dependent spectral function for a J=1/2
system, such as the nucleon, 
it is found that, within the
light-front dynamics with a fixed number of constituents and in the
valence approximation,
only three of the six leading twist T-even transverse-momentum distributions 
are independent.
\end{abstract}

\section{Introduction}
\label{intro}
Information on the three-dimensional proton structure
can be obtained from the quark transverse momentum distributions (TMDs) 
\cite{uno}, which can be
accessed through semi-inclusive deep inelastic electron scattering (SIDIS). 
In  particular the single spin asymmetries (SSAs) allow one to 
extract the Sivers and the Collins 
contributions,  expressed in terms of different TMDs
and {{fragmentation functions}} (ff) \cite{uno}. 
A large Sivers asymmetry 
was measured in ${{{\vec p}(e,e'\pi)x}}$
\cite{due} and a small one in ${{{\vec D}(e,e'\pi)x}}$ \cite{tre},
showing a strong flavor dependence of TMDs.
To clarify the flavour dependence
of the quark transverse momentum distributions, high precision 
experiments, involving both protons and neutrons,
are needed \cite{quattro}.

In Ref. \cite{cinque} an experiment
to extract information on the neutron TMDs from 
experimental measurements of the SSAs on $^3$He, at JLab at 12 GeV, 
was proposed.
To obtain a reliable information one has to take carefully into account 
the structure
of  $^3$He, the interaction in the final state (FSI) 
between the observed pion and the remnant
debris, and the relativistic effects. 
The present paper reports on our efforts about these
items. 
\section{SIDIS off $^3$He}
A polarized $^3$He is an ideal target 
to study the polarized {{neutron}} since, at a 90\% level, 
a polarized $^3$He is equivalent to
a polarized neutron.
Dynamical nuclear effects in inclusive deep inelastic electron scattering (DIS)
 {{$^3\vec{He}( e,e')X$}} (DIS) were evaluated \cite{sei} with a realistic $^3{{\vec{He}}}$
spin-dependent spectral function, ${{ P^{\tau}_{\sigma,\sigma{\prime}} ({\bf p}, E,S_{He})}}$, with  
$\bf p$ the initial nucleon momentum in the laboratory frame and $E$ the missing energy \cite{sette}. 
It was found  
that the formula
\begin{equation}
  {{A_n }}\simeq {1 \over 
{{p_n}} f_n} \left 
( {A^{exp}_3} - 2 
{p_p} f_p ~
{{A^{exp}_p}} \right ) 
\label{formula}
\end{equation}
can be safely adopted to extract the neutron information 
from $^3$He and proton data.
This formula is actually widely used by experimental collaborations.
The nuclear effects are hidden
in the proton and  neutron {{"effective polarizations"}}, $ p_{p(n)}$.
 With the AV18 nucleon-nucleon interaction \cite{otto} 
the values
$ {{p_p}} = -0.023  $, 
${{p_n}}= 0.878 $ were obtained \cite{nove}.
The quantities $f_{p(n)}$ in Eq. (\ref{formula}) are the dilution factors.
To investigate if an analogous formula can be used to extract the {{SSAs}},
in \cite{nove}
the processes {{$^3\vec{He}( e,e'\pi^{\pm})X$}}, with $^3$He 
transversely polarized,
were evaluated 
in the Bjorken limit and
in PWIA, i.e. the  final state interaction (FSI) was considered only 
within the  two-nucleon 
 spectator pair 
 which recoils.
 In such a framework, {{SSAs}} for $^3$He
involve convolutions of 
${{ P^{\tau}_{\sigma,\sigma^{\prime}} ({\bf p}, E,S_{He})}}$,
 with TMDs
and  ffs. 
Ingredients of the calculations were: i)
a realistic 
${{ P^{\tau}_{\sigma,\sigma^{\prime}} ( {\bf p}, E,S_{He})}}$
for {{$^3$He}},
obtained
using the {{AV18}} interaction;
ii) parametrizations of data for TMDs and 
ff, whenever available;
iii) models for the unknown TMDs and 
ff.
It was found that,
in the Bjorken limit, the extraction procedure through the  formula
successful in DIS  works nicely 
for the Sivers and Collins SSAs \cite{nove}.
The generalization of Eq. (\ref{formula}) 
to extract Sivers and Collins asymmetries
from $^3$He and proton asymmetries was recently
used by experimental collaborations \cite{dieci}.

In SIDIS experiments off $^3$He, the relative energy between the spectator $(A-1)$ system
and the system composed by the detected pion and the remnant debris (see Fig. 1) is a few GeV 
and FSI can be treated through a generalized eikonal
approximation (GEA) \cite{undici}. The GEA was already succesfully applied
to unpolarized SIDIS in Ref. \cite{dodici}.
The FSI effects to be considered are due to the propagation of the debris, formed after the
 $\gamma^*$ absorption by a target quark, and the subsequent hadronization, both of them
 influenced by the presence of a fully interacting $(A-1)$ 
spectator system (see Fig. 1).
\begin{figure}[h]
\includegraphics[width=7.5cm]{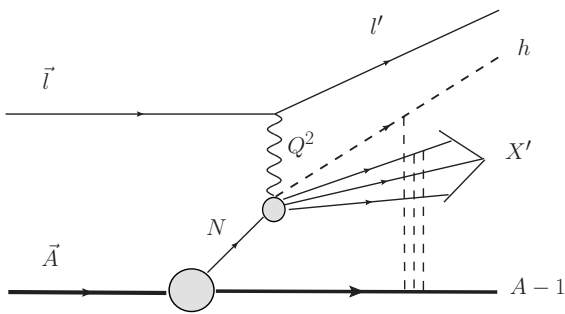}
\caption{Interaction between the $(A-1)$ spectator system and the debris produced by the
absorption of a virtual photon by a nucleon in the nucleus.}
\label{fig:2}
\end{figure}
\hspace{5.cm} 

In this approach, the key quantity to introduce 
the FSI effects is the {\em{distorted}} 
spin-dependent spectral function, whose
relevant part in the evaluation of SSAs is:
\vskip 2mm
\begin{eqnarray}
\quad \quad \quad
{\cal P}_{||}^{PWIA({{FSI}})}=
{\cal O}^{IA({{FSI}})}_{\frac12 \frac12}-
{\cal O}^{IA({{FSI}})}_{-\frac12 -\frac12}
\quad \quad \quad
\end{eqnarray}
with:
\vskip -3mm
\begin{eqnarray}
{\cal O}^
{IA}_{\lambda\lambda'}(\vec p,E) 
= \sum \! \!\! \!\! \!\! \!\int{~d\epsilon^*_{A-1}} ~
\rho\left(\epsilon^*_{A-1}\right)~
\langle   
S_A, 
{{{\bf P}_A}}
| 
\Phi_{\epsilon^*_{A-1}},\lambda',\vec p \rangle
 \langle  
\Phi_{\epsilon^*_{A-1}},
\lambda,\vec p|  S_A,{{{\bf P}_A}}\rangle ~
\delta\left( E- B_A-\epsilon^*_{A-1}\right) ~,
\label{overlap} 
\end{eqnarray}
and
\begin{eqnarray}
{\cal O}^
{{FSI}}_{\lambda\lambda'}(\vec p,E) 
& = & \sum \! \!\! \!\! \!\! \!\int{~d\epsilon^*_{A-1}} ~
\rho\left(\epsilon^*_{A-1}\right)~
\langle   
S_A, 
{{{\bf P}_A}}
| ({{\hat S_{Gl}}})
\{\Phi_{\epsilon^*_{A-1}},\lambda',\vec p\} \rangle
 \langle  
({{\hat S_{Gl}}})
\{\Phi_{\epsilon^*_{A-1}},
\lambda,\vec p\}|  S_A,{{{\bf P}_A}}\rangle 
\nonumber
\\
& \times &
\delta\left( E- B_A-\epsilon^*_{A-1}\right) ~,
\label{overlapfsi} 
\end{eqnarray}
where $\rho\left(\epsilon^*_{A-1}\right)$ is the density of the $(A-1)$-system
states with intrinsic energy $\epsilon^*_{A-1}$, 
and $|  S_A,{{{\bf P}_A}}\rangle$ is the
ground state of the $A$-nucleons nucleus with polarization $S_A$.
${\hat S_{Gl}}$ is the {{Glauber}} operator:
\begin{eqnarray}
\, \, {{\hat S_{Gl}}} 
({\bf r}_1,{\bf r}_2,{\bf r}_3)=
\prod_{i=2,3}\bigl[1-\theta(z_i-z_1)
{{\Gamma}}({\bf b}_1-{\bf b}_i,{ z}_1-{z}_i)
\bigr]
\end{eqnarray}
and $\Gamma({{\bf b}_{1i}},z_{1i})$ the generalized {{profile function}}:
\begin{equation}
\quad 
{{\Gamma({{\bf b}_{1i}},z_{1i})}}\, =
\,\frac{(1-i\,\alpha)\,\,
{{\sigma_{eff}(z_{1i})}}} {4\,\pi\,b_0^2}\,\exp 
\left[-\frac{{\bf b}_{1i}^{2}}{2\,b_0^2}\right]~,
\label{profile}
\end{equation}
where ${\bf r}_{1i}=\{{\bf b}_{1i}, {\bf z}_{1i}\}$
with ${\bf z}_{1i} ={\bf z}_{1}-{\bf
z}_{i}$ and ${\bf b}_{1i}={\bf b}_{1}-{\bf b}_{i}$.
The models for the profile function, $\Gamma({{\bf b}_{1i}},z_{1i})$, and for the 
effective cross section, $\sigma_{eff}(z_{1i})$, as well as the values of the parameters 
$\alpha$ and $b_0$ are the ones 
exploited in Ref. \cite{dodici} to nicely describe the 
JLab data corresponding to the unpolarized spectator SIDIS off the deuteron.  

It occurs that ${\cal P}_{||}^{PWIA}$
and 
${\cal P}_{||}^{FSI}$
can be very different, but the relevant observables for the SSAs
involve integrals, dominated
by the low momentum region, where the  FSI effects on ${\cal P}_{||}$ are 
minimized and the
spectral function is large \cite{undici}.
As a consequence the effective nucleon polarizations change from $ {{p_p}} = -0.023  $, 
${{p_n}}= 0.878 $ to ${{p_p^{FSI}}} = -0.026$, ${{p_n^{FSI}}}= 0.76 $, 
where 
\begin{equation}
  {p^{FSI}_{p(n)}} = 
  \int d\epsilon_{S} \int d{\bf p} ~ 
{ {Tr [{\bf{S}}_{He} * {\bf{\sigma}} ~
  {P}^{p(n)}_{FSI}({\vec p},E,S_{He}) ]
}}~, 
\label{polFSI}
\end{equation}
with  ${ P}^{p(n)}_{FSI}(\vec p,E,S_{He})$  the distorted spin-dependent
spectral function, defined in terms of the overlaps of 
Eq. (\ref{overlapfsi}) \cite{undici}.
Then $p_{p(n)}$ with and without the FSI differ
by 10-15\% . 
Actually, one has also to consider the effect of the FSI on dilution factors. 
We have found,
in a wide range of kinematics typical for the experiments at JLAB 
\cite{cinque,dieci},
that the product
of polarizations and dilution factors changes very little \cite{tredici}.
Indeed the effects of FSI in the dilution factors and in the
effective polarizations are found to compensate each other
to a large extent and the {{usual extraction}} appears to be safe :
\begin{eqnarray}
  {{A_n }}\simeq {1 \over {{p_n^{FSI}}} f_n^{FSI}} 
  \left ( {A^{exp}_3} - 2 {p_p^{FSI} f_p^{FSI} {{A^{exp}_p}}} \right )~ \simeq
{1 \over {{p_n}} f_n }
  \left ( {A^{exp}_3} - 2 {p_p f_p ~ {{A^{exp}_p}}} \right )~ \quad   
\vspace{-0.1cm}
\end{eqnarray}
In \cite{nove} the calculation was performed within a non relativistic approach for the spectral function,
but with the correct relativistic kinematics in the Bjorken limit.
For an accurate description of SIDIS processes, the role played by {{relativity}} 
has to be fully investigated: it will become even more important with the upgrade
of JLab @ 12 GeV. 
To study relativistic effects in the 
actual experimental kinematics, we adopted \cite{14} the
Relativistic Hamiltonian Dynamics (RHD) introduced by 
Dirac \cite{15}.
Indeed the {{RHD} of an interacting system with a fixed number of on-mass-shell constituents, 
with the Bakamijan-Thomas 
construction of the Poincar\'e generators \cite{16},
is fully Poincar\'e covariant.
The  Light-Front (LF) form of RHD has a {{subgroup}} structure of
the {{LF boosts}},
allows a separation of the {{intrinsic motion}} from
the global one, which is very important for the description of DIS, SIDIS and 
 deeply virtual Compton scattering (DVCS) processes,
and allows a {{meaningful Fock expansion}}. 
The key quantity to consider in SIDIS processes is the LF relativistic spectral function,
$
{\cal P}^{\tau}_{\sigma'\sigma}(\tilde{\bf \kappa},\epsilon_S,S_{He})
$, with $\tilde{\bf \kappa}$ an intrinsic nucleon momentum and $\epsilon_S$ 
the energy of the two-nucleon
 spectator  system. 
The  LF  spectral function will be very useful also for other studies
(e.g., for nuclear generalized parton distributions (GPDs), where final states
have to be properly boosted, studied so far only within a non-relativistic
spectral function \cite{17,18}).
The LF nuclear spectral function,
$
{\cal P}^{\tau}_{\sigma'\sigma}(\tilde{\bf \kappa},\epsilon_S,S_{He})
$,
 is defined in terms of LF overlaps \cite{19} between the  ground state of a polarized $^3$He and
 the cartesian product of an interacting state of two nucleons with energy $\epsilon_S$
 and a plane wave for the third
 nucleon. Within a reliable approximation \cite{19} it can be
 given in terms of the unitary Melosh Rotations, 
$
{{D^{{1 \over 2}} [{\cal R}_M ({\blf \kappa})]}}$,
and the usual {{instant-form spectral function}}
$
{{
{P}^{\tau}_{\sigma'_1\sigma_1}
}}$:
\bq
{ {
  {\cal P}^{\tau}_{\sigma'\sigma}({\blf \kappa},\epsilon_{S},S_{He})
}}
\propto  
~\sum_{\sigma_1 \sigma'_1} 
{{D^{{1 \over 2}} [{\cal R}_M^\dagger ({\blf
\kappa})]_{\sigma'\sigma'_1}}}~
{{
{P}^{\tau}_{\sigma'_1\sigma_1}({\bf p},\epsilon_{S},S_{He})
}} ~
{{D^{{1 \over 2}} [{\cal R}_M ({\blf \kappa})]_{\sigma_1\sigma}}}  
\eq
 \vskip -0.2cm
 
\begin{table*}
\centering
\caption{Proton and neutron effective polarizations within the non relativistic appproach (NR)
and preliminary results within the light-front relativistic dynamics approach (LF). First line : 
longitudinal effective
polarizations; second line : transverse effective
polarizations.}
\label{tab-1}
\begin{tabular}{|c|c|c|c|c|c|}
\hline  &\hspace{-1mm}$proton \, {NR}$\hspace{-2mm}&\hspace{-1mm}$proton \,
{LF}$\hspace{-2mm}&\hspace{-1mm}$neutron \, {NR}$\hspace{-2mm}&\hspace{-1mm}$neutron \, {LF}$\hspace{-1mm} \\ 
\hline 
$\int^{~}_{~} dE d\vec{p}\,
~ Tr( {\cal{P}} \sigma_{z})_{\vec{S}_A=
\widehat{z}}$ 
\hspace{3mm} 
& -0.02263 & {{-0.02231}} & 0.87805 & 
{{0.87248}} \\ 
\hline 
\hspace{-7mm} $\int^{~}_{~} dE d\vec{p}\,
~ Tr( {\cal{P}} \sigma_{y})_{\vec{S}_A=
\widehat{y}}$ \hspace{-2mm}
& -0.02263 & {{-0.02268}} & 0.87805 & {{0.87494}} \\ 
\hline 
\end{tabular} 
\end{table*}
  Notice that the instant-form spectral funtion
$
{P}^{\tau}_{\sigma'_1\sigma_1}
({\bf p},\epsilon_{S},S_{He})
$
is given in terms of {{three independent functions}},
${{B_{0},B_{1},B_{2}}}$ \cite{sette},
once parity and t-reversal are imposed:
 \bq
\hspace{-0.8cm}{{
{P}^{\tau}_{\sigma'_1\sigma_1}({\bf p},\epsilon_{S},S_{He})
}}
= 
\left[ 
{{
B_{0,S_{He}}^{\tau}\hspace{-0.1cm}(|{\bf p}|,E)
}}
+
{\bm \sigma} \cdot {\bf f}^{\tau}_{S_{He}}\hspace{-0.1cm}
({\bf p},E) \right]_{\sigma'_1\sigma_1}\hspace{-.3cm}
\eq
with 
\bq
\hspace{-.9cm}
{\bf f}^{\tau}_{S_{He}}({\bf p},E) =
{\bf S}_{He}
{{
B_{1,S_{He}}^{\tau}\hspace{-0.1cm}(|{\bf p}|,E)}}+{\bf\hat{p}}~{\bf(\hat{p}}
\cdot {\bf S}_{He}) 
{{
B_{2,S_{He}}^{\tau}\hspace{-0.1cm}(|{\bf p}|,E)
}}\hspace{-.3cm}
\eq
Adding FSI, more terms should be included.

We are now  evaluating the SSAs using the {{LF hadronic tensor}},
at finite values of $Q^2$. 
%
The preliminary results are quite encouraging, since, as shown in Table 1,
{{LF}} longitudinal and transverse polarizations change little from the ones 
obtained within the NR spectral function and weakly differ from
each other. 
Then we find that
the extraction procedure works well within 
{{the LF approach}} as it does in the non
relativistic case.

Concerning the FSI, we plan to include in our LF approach the FSI
between the jet produced from the hadronizing quark
 and the two-nucleon system through the Glauber approach of 
Ref. \cite{undici}.

\section{The $J =1/2$ {{LF}} spectral function and the nucleon {{LF}} TMDs}

The TMDs for a nucleon with total momentum $P$ and spin $S$ are introduced
through the {{q-q correlator}}
\begin{eqnarray}
{{\Phi(k, P, S)}}_{\alpha\beta} 
& = &
\int {d^4z} ~ e^{i k{\cdot}z}
\langle P S | ~\bar \psi_{q\beta}(0) ~ \psi_{q\alpha} (z) | P S \rangle 
= \frac12
\{ \phantom{\frac1M} \hspace{-4mm} 
    {{A_1}} 
\, {P}\hspace{-2mm} / \hspace{1mm} +
    { {A_{2}}} 
\, S_L \, \gamma_5 \, {P}\hspace{-2mm} / \hspace{1mm} +  { {A_3}} 
\, {P}\hspace{-2mm} / \, \gamma_5 \, {S}_\perp\hspace{-4mm}  
\nonumber 
\\
& + &   
\frac1{M} \, 
{{\widetilde{A}_1}} 
\, \vec{k}_\perp{\cdot}\vec{S}_\perp \,
    \gamma_5 {P}\hspace{-2mm} /  \hspace{1mm}  + \,
{{\widetilde{A}_2}} 
\, \frac{S_L}{M} \,
    {P}\hspace{-2mm} / \, \gamma_5 \, {k}_\perp\hspace{-4mm} 
+  \, \frac1{M^2} \, 
{{\widetilde{A}_3}} 
\, \vec{k}_\perp{\cdot}\vec{S}_\perp \,
    {P}\hspace{-2mm} / \, \gamma_5 \, {k}_\perp\hspace{-4mm} / \hspace{2mm}
\}_{\alpha\beta} ,~~
\end{eqnarray}
where $k$ is the parton momentum in the laboratory frame,
so that the {{six twist-2 T-even functions}}, {{$A_i,~
\widetilde{A}_i ~ (i=1,3)$}}, can be obtained by proper traces of 
{{$\Phi(k, P, S)$}}.

Indeed, particular combinations of the  functions 
{{$A_i,\widetilde{A}_i ~ (i=1,3)$ }} 
can be obtained by the following traces of 
{{$\Phi(k, P, S)$}} :
\vskip -4mm
\begin{eqnarray}
   \frac1{2P^+} \, {{\Tr(\gamma^+ \Phi)}}
 & = &
{{ A_1}} 
\,,
\nonumber
\\
  \frac1{2P^+} \, {{\Tr(\gamma^+ \gamma_5 \Phi)}}   & =  &
  S_L \, 
{{  A_{2}} ~
+ ~
  \frac1{M} \, \vec{k}_\perp{\cdot}\vec{S}_\perp \, 
{{\widetilde{A}_1}}}
\, ,
\nonumber
\\
  \frac1{2P^+} \, {{\Tr(i \sigma^{j+} \gamma_5 \Phi)}} & =  &  S_\perp^j \, 
{{A_3}} ~ +
~ \frac{S_L}{M}\, k_\perp^j \, 
{{\widetilde{A}_2}} ~ + 
  \frac1{M^2} ~   \vec{k}_\perp{\cdot}\vec{S}_\perp ~ k_\perp^j ~ 
{{\widetilde{A}_3}}  
\, ~~ (j=x,y).  
\label{traces}
\end{eqnarray}

\vskip 2mm
The  {six {T-even  twist-2 TMDs}} 
 for the quarks inside a nucleon can be obtained by integration
of the functions {{$~ A_i, ~ \widetilde{A}_i $}} ~ on ~ {{$k^+$}} ~ and ~ {{$k^-$}} as follows
\bq
\hspace{-.6cm}f(x, |{\bf k}_{\perp}|^2 )
 = \frac1{2} \int \frac{d{ k^+} d{ k^-}}{(2\pi)^4} ~ \delta[k^+ - x P^+] ~ 2P^+  A_1 \, , \quad
 \nonumber\\ 
\Delta f(x, |{\bf k}_{\perp}|^2 )
 = \frac1{2} \int \frac{d{ k^+} d{ k^-}}{(2\pi)^4} ~ \delta[k^+ - x P^+] ~ 2P^+  A_2 \, , 
 \nonumber\\ 
 g_{1T}(x, |{\bf k}_{\perp}|^2 )
= \frac1{2} \int \frac{d{ k^+} d{ k^-}}{(2\pi)^4} ~ \delta[k^+ - x P^+] ~ 2P^+ \widetilde{A}_1 \, , 
\nonumber \\
\Delta'_T   f(x, |{\bf k}_{\perp}|^2 )
= \frac1{2} \int \frac{d{ k^+} d{ k^-}}{(2\pi)^4} ~ \delta[k^+ - x P^+] ~ 2P^+ 
\left ({A}_3 + {|{\bf k}_{\perp}|^2 \over 2M^2} \widetilde{A}_3\right ) \, , \nonumber \\ 
h^{\perp}_{1L}(x, |{\bf k}_{\perp}|^2 )
= \frac1{2} \int \frac{d{ k^+} d{ k^-}}{(2\pi)^4} ~ \delta[k^+ - x P^+] ~ 2P^+ \widetilde{A}_2 \, , 
\nonumber 
\eq
\bq
\hspace{-.6cm}h^{\perp}_{1T}(x, |{\bf k}_{\perp}|^2 )
   = \frac1{2} \int \frac{d{ k^+} d{ k^-}}{(2\pi)^4} ~ \delta[k^+ - x P^+] ~ 2P^+  \widetilde{A}_3~.
\eq

Let us consider the  contribution to the {{correlation function}}
from  on-mass-shell fermions
\vskip -0.3cm
\bq
\hspace{-.8cm}{ {~\Phi_p(k,P,S)=~{(~{  k \hspace{-2mm} /}_{on}~ + ~m )\over 2 m}~
{{\Phi(k,P,S)}}~{(~{  k\hspace{-2mm} /}_{on}~ + ~m )\over 2 m} }}= \hspace{-0.2cm}
\label{PC}
\eq 
\bq \hspace{-0.8cm} = 
\sum_{\sigma\sigma'}~u_{LF}({\tilde k},\sigma')~\bar{u}_{LF}( {\tilde k},\sigma')
~{{\Phi(k,P,S)}}
~u_{LF}({\tilde k},\sigma)\bar{u}_{LF}( {\tilde k},\sigma)~,
\hspace{-0.3mm}  \nonumber
\eq 
and let us identify $\bar{u}_{LF}( {\tilde k},\sigma')
~\Phi(k,P,S)
~u_{LF}({\tilde k},\sigma)$, up to a kinematical factor $K$,
with 
the {{LF nucleon spectral function}},
${\cal P}^{}_{\sigma'\sigma}(\tilde{\bf \kappa},\epsilon_S,S)$ \cite{19}:
\bq
\hspace{-0.5cm}{{\bar{u}_{LF}( {\tilde k},\sigma')
~\Phi(k,P,S)
~u_{LF}({\tilde k},\sigma)}} = {{K}} ~ 
{{{\cal P}^{}_{\sigma'\sigma}(\tilde{\bf \kappa},\epsilon_S,S)}}~.
\label{id}
\eq
In a reference frame where ${\bf P}_\perp = 0$, the following relation holds
between the off-mass-shell minus component {$k^-$} of the momentum of the struck quark and 
the spectator diquark energy {{$\epsilon_S $}}  :
\vskip -0.3cm
\bq
{{k^-}} = {{M^2}\over {P^+}} ~ - ~ {{({{\epsilon_S}} + m) ~ 4m + |{\bm k}_\perp|^2}\over {P^+ -k^+}}~.
\eq
Let us approximate the full correlation function ${{\Phi(k,P,S)}}$ by its particle contribution
${{~\Phi_p(k,P,S)}}$. Then, through relations analogous to the ones of Eq. (\ref{traces}),
which allow one to obtain the functions $A_i,~
\widetilde{A}_i$
 from the traces of 
${{\Phi(k,P,S)}}$, 
 the valence approximations 
${{~ A_i^V, ~ \widetilde{A}_i^V ~ (i=1,3)}}$ 
for the functions ${{~ A_i, ~ \widetilde{A}_i }}$  can be obtained
by the  
 traces  $~[\gamma^+ ~\Phi_p(k,P,S)]$,~ $[\gamma^+ ~\gamma_5~~\Phi_p(k,P,S)]$, and
 $[{k}\hspace{-2mm} / _\perp\gamma^+
\gamma_5~\Phi_p(k,P,S)]$.  
However these same traces can be also expressed through the {{LF spectral function}},
since from Eqs. (\ref{PC},\ref{id}) one has
\bq
\hspace{-.7cm}{ {~ Tr\left[ \gamma^+ ~\Phi_p(k,P,S) \right]}} ~ 
=
~{k^+\over m } ~ K ~ { {Tr\left[ {\cal P}^{}_{}(\tilde{\bf \kappa},\epsilon_S,S)
\right]}}
\eq
\vspace{-6mm}
\bq
\hspace{-.9cm}{ {~Tr\left[ \gamma^+ ~\gamma_5~~\Phi_p(k,P,S)\right]}}
=
 {k^+\over {m }} ~ K ~ { {
Tr\left[ \sigma_z ~ {\cal P}^{}(\tilde{\bf \kappa},\epsilon_S,S) \right]}}
\eq
\vspace{-6mm}
\bq
\hspace{-0.9cm}{ {
Tr\left[ {k}\hspace{-2mm} / _\perp\gamma^+
\gamma_5~\Phi_p(k,P,S)\right]}} 
={k^+\over 
m
}  K  {{
Tr\left[ {\bm k}_\perp \cdot {\bm \sigma} ~ 
{\cal P}^{}(\tilde{\bf \kappa},\epsilon_S,S) \right]}} \hspace{-.4cm}~
\eq
In turn, as in the
$^3$He case, the traces 
${\Tr}( {\cal P} I)$, 
${\Tr}( {\cal P} \sigma_z )$, 
$
{\Tr}(  {\cal P} \sigma_i)$ ($i=x,y$) can be expressed in terms of three scalar functions, 
$B_0$, $B_1$, $B_2$
and known
kinematical factors.
Then, within the {{LF approach}} with a fixed number of particles and in the valence approximations, 
the six 
leading twist T-even functions  
${{~ A_i^V, ~ \widetilde{A}_i^V ~ (i=1,3)}}$ 
can be expressed in terms of the  three independent scalar functions 
${ {B_0, B_1, B_2}}$.
\\
Therefore only three 
of the six T-even TMDs are independent \cite{19}.
We stress however that this result, not valid in general in QCD
(see, e.g., Ref. \cite{pasqui}),
is a prediction of our peculiar framework, i.e.,
hamiltonian light-front dynamics with a fixed number
of constituents, 
finalized to yield a proper Poincar\'e covariant description of the
nucleon in the valence approximation.
Its experimental check would be
therefore a test of the reliability of our scheme to describe SIDIS
processes in the valence region.
\vspace{-2mm}
\section{Conclusions}
A realistic study of the DIS processes off $^3\vec{He}$ and in particular
of the SSAs in the reaction {{ $^3\vec{He}(e,e'\pi^{\pm})X$ }} beyond the 
PWIA and the non relativistic
approach
is under way.
The {{{{FSI effects}}}} have been evaluated through the  
{{GEA}} and the introduction of
a {{distorted spin-dependent spectral
function}}. 
The
{{{{relativistic effects}}}} are studied
through the analysis of a {LF} spectral function
(up to now only in PWIA).
Preliminary results are encouraging,
in view of a sound extraction of the neutron
information from experimental data. 
Nuclear effects
in the extraction of the {{neutron}} information are 
found to be {{under control}}, even when the interaction 
in the final state is considered,
and the relativistic effects appear to be small.
An analysis at finite $Q^2$  with the {LF} spectral function
is in progress.
The next step is to
complete this program.  
Then we will
apply the {LF} spectral function to other processes
(e.g., DVCS) to exploit other possibilities to use
{{{{$^3$He as an effective neutron target and as 
a laboratory for 
light-front
studies}}}}. 

The introduction of a LF spin-dependent spectral function for a nucleon 
allowed us to find
relations  among the six leading twist T-even TMDs,
which are exact 
{{within LF dynamics with a fixed number of degrees of freedom,
in the valence approximation}}.
It can be shown that, in this case, {{only three}}
of the six {{T-even TMDs are independent}}.
The above relations, although not true in QCD,
could be experimentally
checked to test our {{LF}} description of SIDIS in the valence region.
%
%
%

\end{document}